\documentclass[sigconf]{acmart}
\usepackage{booktabs}
\usepackage{tikz}
\usetikzlibrary{external}
\usepackage{ textcomp }
\usepackage[utf8]{inputenc}
\usepackage{enumitem}
\newcommand{\xsub}[1]{%
  \mbox{\scriptsize\begin{tabular}{@{}c@{}}#1\end{tabular}}%
}

\usepackage{makecell}
\usepackage{threeparttable}
\usepackage{longtable}
\usepackage{multirow}
\usepackage{tabularx}
\usepackage{tabulary}
\usepackage{array}
\newcommand{\PreserveBackslash}[1]{\let\temp=\\#1\let\\=\temp}

\newcolumntype{C}[1]{>{\PreserveBackslash\centering}p{#1}}
\newcolumntype{R}[1]{>{\PreserveBackslash\raggedleft}p{#1}}
\newcolumntype{L}[1]{>{\PreserveBackslash\raggedright}p{#1}}
\usepackage{amssymb}
\usepackage{eqnarray}
\usepackage{amsmath}
\usepackage{epstopdf}
\usepackage{graphicx}
\usepackage{subfig}
\usepackage{caption}
\usepackage{color}
\usepackage{url}
\usepackage{hyperref}
\usepackage{bm}
\usepackage{balance}
\usepackage[linesnumbered,ruled,vlined]{algorithm2e}

\SetCommentSty{mycommfont}

\setcopyright{rightsretained}

\begin{document}

\title{Aspect-Aware Latent Factor Model: \\ Rating Prediction with Ratings and Reviews}

\author{Zhiyong Cheng}
\affiliation{%
	\institution{National University of Singapore}
}
\email{jason.zy.cheng@gmail.com}

\author{Ying Ding}
\affiliation{%
	\institution{	Vipshop Inc., USA}
}
\email{ian.yingding@gmail.com}

\author{Lei Zhu}
\affiliation{%
	\institution{Shandong Normal University, China}
}
\email{leizhu0608@gmail.com}

\author{Mohan Kankanhalli}
\affiliation{%
	\institution{National University of Singapore}
}
\email{mohan@comp.nus.edu.sg}
\fancyhead{}

\begin{abstract}
Although latent factor models (e.g., matrix factorization) achieve good accuracy in rating prediction, they suffer from several problems including cold-start, non-transparency, and suboptimal recommendation for local users or items. In this paper, we employ textual review information with ratings to tackle these limitations. Firstly, we apply a proposed aspect-aware topic model (ATM) on the review text to model user preferences and item features from different \emph{aspects}, and estimate the \emph{aspect importance} of a user towards an item. The aspect importance is then integrated into a novel aspect-aware latent factor model (ALFM), which learns user's and item's latent factors based on ratings. In particular, ALFM introduces a weighted matrix to associate those latent factors with the same set of aspects discovered by ATM, such that the latent factors could be used to estimate aspect ratings. Finally, the overall rating is computed via a linear combination of the aspect ratings, which are weighted by the corresponding aspect importance. To this end, our model could alleviate the data sparsity problem and gain good interpretability for recommendation. Besides, an aspect rating is weighted by an aspect importance, which is dependent on the targeted user's preferences and targeted item's features. Therefore, it is expected that the proposed method can model a user's preferences on an item more accurately for each user-item pair locally. Comprehensive experimental studies have been conducted on 19 datasets from Amazon and  Yelp 2017 Challenge dataset. Results show that our method achieves significant improvement compared with strong baseline methods, especially for users with only few ratings. Moreover, our model could interpret the recommendation results in depth.

\end{abstract}

\copyrightyear{2018}
\acmYear{2018} 
\setcopyright{iw3c2w3}
\acmConference[WWW 2018]{The 2018 Web Conference}{April 23--27, 2018}{Lyon, France}
\acmBooktitle{WWW 2018: The 2018 Web Conference, April 23--27, 2018, Lyon, France}
\acmPrice{}
\acmDOI{10.1145/3178876.3186145}
\acmISBN{978-1-4503-5639-8/18/04}



\maketitle

\section{Introduction}
When making comments on an item (e.g., \emph{product}, \emph{movie}, and \emph{restaurant}) in the online review/business websites, such as  Yelp and Amazon, reviewers also provide an overall rating, which indicates their overall preference or satisfaction towards the corresponding items. Hence, predicting users' overall ratings to unrated items or \emph{personalized rating prediction} is an important research problem in recommender systems. Latent factor models (e.g., matrix factorization~\cite{koren2009matrix,zhang2016discrete,cheng2017exploiting}) are the most widely used and successful techniques for rating prediction, as demonstrated by the Netflix Prize contest~\cite{bell2007lessons}. These methods characterize user's interests and item's features using \emph{latent factors} inferred from rating patterns in user-item rating records.
As a typical collaborative filtering technique, the performance of MF suffers when the ratings of items or users are insufficient ( also known as the cold-start problem)~\cite{he2015trirank}. Besides, a rating only indicates the overall satisfaction of a user towards an item, it cannot explain the underlying rationale. For example, a user could give a restaurant a high rating because of its delicious food or due to its nice ambience. Most existing MF models cannot provide such fine-grained analysis.  Therefore, relying solely on ratings makes these methods hard to explicitly and accurately model users' preferences~\cite{wang2018rec,he2015trirank,ling2014ratings,mcauley2013hidden,wu2015flame}.

Moreover, MF cannot achieve optimal rating prediction locally for each user-item pair, because it learns the latent factors of users ($\bm{p_u}$) and items ($\bm{q_i}$) via a global optimization strategy~\cite{christakopoulou2016local}. In other words,  $\bm{p_u}$ and $\bm{q_i}$ are optimized to achieve a global optimization over all the user-item ratings in the training dataset.\footnote{In the paper, unless otherwise specified, notations in bold style denote matrices or vectors, and the ones in normal style denote scalars.} As a result, the performance could be severely compromised locally for individual users or items.  MF predicts an unknown rating by the dot product of the targeted user $u$'s and item $i$'s latent factors (e.g., $\bm{p_u}^T\bm{q_i}$). The overall rating of a user towards an item ($\hat{r}_{u,i}$) is decided by the importance/contribution of all factors. Take the $k$-th factor as an example, its contribution  is $p_{u,k}*q_{i,k}$. For accurate prediction, it is important to accurately capture the importance of each latent factor for a user towards an item.  It is well-known that different users may care about different \emph{aspects} of an item. For example, in the domain of restaurants,  some users care more about the taste of \emph{food } while others pay more attention to the \emph{ambience}. Even for the same aspect,  the preference of users could be different from each other. For example, in the \emph{food} aspect, some users like \emph{Chinese cuisines} while some others favor \emph{Italian cuisines}. Similarly, the characteristics of  items on an aspect could also be different from each other. Thus, it is possible that ``a user $u$ prefers item $i$ but dislikes item $j$  on a specific aspect", while ``another user $u'$ favors item $j$ more than item $i$ on this aspect". Therefore, in MF, the importance of a latent factor for users towards an item should be treated differently. At first glance, MF achieves the goal as the influence of a factor (e.g., $k$-th factor) is dependent on both $p_{u,k}$ and $q_{i,k}$ (i.e., $p_{u,k}*q_{i,k}$). However, it is suboptimal to model the importance of a factor by a fixed value of an item or a user.  In fact, MF treats each factor of an item with the same importance to all users (i.e., $q_{i,k}$); and similarly, each factor of a user is equally important to all items (i.e., $p_{u,k}$) in rating prediction. Take the previous example, ``\emph{a user $u$ prefers item $i$ but dislikes item $j$ on an aspect}", i.e., a factor (e..g, $k$) in MF), which means $p_{u,k}*q_{i,k}$ should be larger than $p_{u,k}*q_{j,k}$ (i.e., $p_{u,k}*q_{i,k}>p_{u,k}*q_{j,k}$). On the other hand, ``\emph{user $u'$ favors item $j$ more than item $i$ on this aspect}", thus $p_{u',k}*q_{j,k}$ should be larger than $p_{u',k}*q_{i,k}$ (i.e., $p_{u',k}*q_{i,k} <p_{u',k}*q_{j,k}$). Because the values of $p_{u,k}$ and $p_{u',k}$ are kept the same when predicting ratings, it is impossible for MF to satisfy the local requirements $p_{u,k}*q_{i,k}>p_{u,k}*q_{j,k}$ and  $p_{u',k}*q_{i,k} <p_{u',k}*q_{j,k}$ simultaneously for these user-item pairs. A straightforward solution is to assign different weights (e.g., $w_{u,i,k}$) to different user-item pairs (e.g., $p_{u,k}*q_{i,k}$). However, how to compute a proper weight for each user-item pair is challenging. 

A large amount of research effort has been devoted to deal with these weaknesses of MF methods. For example, various types of side information have been incorporated into MF to alleviate the cold-start problem, such as tags~\cite{shi2013mining,zhang2014attribute}, social relations~\cite{ma2011recommender,wang2017item}, reviews~\cite{ling2014ratings,mcauley2013hidden,zhang2016integrating}, and visual features~\cite{he2016vbpr}. Among them, the accompanied review of a rating contains important complementary information. It not only encodes the information of user preferences and item features but also explains the underlying reasons for the rating. Therefore, in recent years, many models have been developed to exploit reviews with ratings to tackle the cold-start problem and also enhance the explainability of MF, such as  HFT~\cite{mcauley2013hidden}, CTR~\cite{wang2011collaborative},  RMR~\cite{ling2014ratings}, and RBLT~\cite{tan2016rating}. However, a limitation of these models is that they all assume an \emph{one-to-one correspondence relationship} between latent topics (learned from reviews) and latent factors (learned from ratings), which not only limits their flexibility on modeling reviews and ratings but also may not be optimal. In addition, they cannot deal with the suboptimal recommendation for local users or items in MF. In fact, very few studies  in literature have considered this problem.

In this paper, we focus on the problem of \emph{personalized rating prediction} and attempt to tackle the above limitations together by utilizing reviews with ratings. Specifically,  an \emph{A}spect-aware \emph{T}opic \emph{M}odel (ATM) is proposed to extract \emph{latent topics} from reviews, which are used to model users' preferences and items' features in different \emph{aspects}. In particular, each \emph{aspect} of users/items is represented as a probability distribution of latent topics. Based on the results, the relative importance of an aspect (i.e., \emph{aspect importance}) for a user towards an item can be computed. Subsequently, the aspect importance is integrated into a developed \emph{A}spect-aware \emph{L}atent \emph{F}actor \emph{M}odel (ALFM)  to estimate  \emph{aspect ratings}. In particular, a weight matrix is introduced in ALFM to associate the latent factors to the same set of aspects discovered by ATM. In this way, our model avoids referring to external sentiment analysis tools for aspect rating prediction as in~\cite{zhang2014explicit,diao2014jointly}. The overall rating is obtained by a linear combination of the \emph{aspect ratings}, which are weighted by the importance of corresponding aspects (i.e., \emph{aspect importance}).
Note that the latent topics and latent factors in our model are not linked directly; instead, they are correlated via the \emph{aspects} indirectly. Therefore, the number of latent topics and latent factors could be different and separately optimized to model reviews and ratings respectively,  which is fundamentally different from the \emph{one-to-one} mapping in previous models~\cite{mcauley2013hidden,wang2011collaborative,ling2014ratings,tan2016rating,bao2014topicmf,zhang2016integrating}. Besides, our model could learn an aspect importance for each user-item pair, namely, assigning a different weight to each $p_{u,k}*q_{i,k}$, and thus could alleviate the suboptimal local recommendation problem and achieve better performance.

A set of experimental studies has been conducted on 19 real-world datasets from Yelp and Amazon to validate the effectiveness of our proposed model. Experimental results show that  our model significantly outperforms the state-of-the-art methods which also use both reviews and ratings for rating prediction. Besides, our model also obtains better results for users with few ratings, demonstrating the advantages of our model on alleviating the cold-start problem. Furthermore, we illustrate the interpretability of our model on recommendation results with examples. In summary, the main contributions of this work include:

\begin{itemize} [align=left,style=nextline,leftmargin=*,labelsep=\parindent,font=\normalfont]
	\item We propose a novel aspect-aware latent factor model, which could effectively combine reviews and ratings for rating prediction. Particularly, our model relaxes the constraint of one-to-one mappings between the latent topics and latent factors in previous models and thus could achieve better performance.
	
	\item Our model could automatically extract explainable aspects, and learn the aspect importance/weights for different user-item pairs. By associating latent factors with aspects, the aspect weights are integrated with latent factors  for rating prediction. Thus, the proposed model could alleviate the suboptimal problem of MF for individual user-item pairs.
	
	\item We conduct comprehensive experimental studies to evaluate the effectiveness of our model. Results show that  our model is significantly better than previous approaches on tasks of rating prediction,  recommendation for sparse data, and recommendation interpretability.
\end{itemize}


\section{Related Work} \label{sec:relwork}
A comprehensive review on the recommender system is beyond the scope of this work. We mainly discuss the works which utilize both reviews and ratings for rating prediction. Some works assume that the review is available when predicting the rating score, such as SUIT~\cite{li2014suit}, LARAM~\cite{wang2011latent}, and recent DeepCoNN~\cite{zheng2017joint}.  However, in real world recommendation settings, the task should be predicting ratings for the uncommented and unrated items. Therefore, the review is unavailable when predicting ratings. 
We broadly classify the approaches for the targeted task in three categories: (1) sentiment-based, (2) topic-based, and (3) deep learning-based. Our approach falls into the second category.

\textbf{Sentiment-based.} These works analyze user's sentiments on items in reviews to boost the rating prediction performance, such as~\cite{pappas2013sentiment,pero2013opinion,diao2014jointly,zhang2014explicit}. For example, ~\cite{pappas2013sentiment} estimated a sentiment score for each review to build a user-item sentiment matrix, then a traditional collaborative filtering method was applied. Zhang et al.~\cite{zhang2014explicit} analyzed the sentiment polarities of reviews and then jointly factorize the user–item rating matrix. These methods rely on the performance of external NLP tools for sentiment analysis and thus are not self-contained.

\textbf{Topic-based.} These approaches extract latent topics or aspects from reviews. An early work~\cite{ganu2009beyond} in this direction relied on domain knowledge to manually label reviews into different aspects, which requires expensive domain knowledge and high labor cost. Later on, most works attempt to extract latent topics or aspects from reviews automatically~\cite{mcauley2013hidden,bao2014topicmf,diao2014jointly,he2015trirank,ling2014ratings,mcauley2013hidden,wu2015flame,zhang2016integrating,tan2016rating}. A general approach of these methods is to extract latent topics from reviews using topic models~\cite{wang2011collaborative,mcauley2013hidden,ling2014ratings,zhang2016integrating,tan2016rating} or non-negative MF~\cite{bao2014topicmf,qiu2016aspect} and learn latent factors from ratings using MF methods. HFT~\cite{mcauley2013hidden} and  TopicMF~\cite{bao2014topicmf} link the latent topics and latent factors by using a defined transform function.    ITLFM~\cite{zhang2016integrating} and RBLT~\cite{tan2016rating} assume that the latent topics and latent factors are in the same space, and linearly combine them to form the latent representations for users and items to model the ratings in MF. CTR~\cite{wang2011collaborative} assumes that the latent factors of items depend on the latent topic distributions of their text, and adds a latent variable to offset the topic distributions of items when modeling the ratings. RMR~\cite{ling2014ratings} also learns item's features using topic models on reviews, while it models ratings using a mixture of Gaussian rather than MF methods. Diao et al.~\cite{diao2014jointly} propose an integrated graphical model called JMARS to jointly model aspects, ratings and sentiments for movie rating prediction. Those models all assume an one-to-one mapping between the learned latent topics from reviews and latent factors from ratings.  Although we adopt the same strategy to extract latent topics and learn latent factors, our model does not have the constraint of one-to-one mapping. Besides, Zhang et al.~\cite{zhang2014explicit} extracted aspects by decomposing the user–item rating matrix into item–aspect and user–aspect matrices.   He et al.~\cite{he2015trirank}  extracted latent topics from reviews by modeling the user-item-aspect relation with a tripartite graph.  


\textbf{Deep learning-based}. Recently, there has been a trend of applying deep learning techniques in recommendation~\cite{he2017neural,covington2016deep}. For example, He et al. generalized  matrix factorization and factorization machines to  neural collaborative filtering and achieved promising performance~\cite{he2017neural,he2017fm}. Textual reviews have also been used in deep learning models for recommendation~\cite{zhang2016collaborative,zheng2017joint,catherine2017transnets,zhang2017joint}. The most related works in this direction are DeepCoNN~\cite{zheng2017joint} and TransNet~\cite{catherine2017transnets}, which apply deep techniques to reviews for rating prediction. In DeepCoNN, reviews are first processed by two CNNs to learn user's and item's representations, which are then concatenated and passed into a regression layer for rating prediction. A limitation of DeepCoNN is that it uses reviews in the testing phase. ~\cite{catherine2017transnets} shows that the performance of DeepCoNN decreases greatly when reviews are unavailable in the testing phase. To deal with the problem,  TransNet~\cite{catherine2017transnets} extends DeepCoNN  by introducing an additional layer to simulate the review corresponding to the target user-item pair. The generated review is then used for rating prediction.

\begin{table}[]
	\footnotesize
	\centering
	\caption{Notations and their definitions}
	\vspace{-0.2cm}
	\begin{tabular}{ll}
		\Xhline{1.2pt}\noalign{\smallskip}
		Notation & Definition \\
		\noalign{\smallskip}\hline\noalign{\smallskip}
		$\mathcal{D}$ & corpus with reviews and ratings \\
		$d_{u,i}$ &  review document of user $u$ to item $i$ \\
		$s$ & a sentence in a review $d_{u,i}$ \\
		$\mathcal{U}$, $\mathcal{I}$, $\mathcal{A}$ &  user set, item set, and aspect set, respectively \\
		$M,N,A$ & number of users, items, and aspects, respectively\\
		$N_{w,s}$ & number of words in a sentence $s$ \\
		\noalign{\smallskip}\hline\noalign{\smallskip}
		$K$ &  number of latent topics in ATM  \\
		$y$ & an indicator variable in ATM \\
		$a_s$ & assigned aspect $a$ to sentence $s$ \\
		$\pi_u$ & the parameter of Bernoulli distribution $P(y=0)$  \\
		$\eta$ & Beta priors ($\eta=\{\eta_0, \eta_1\}$) \\
		$\bm{\alpha_u}, \bm{\alpha_i}$ & Dirichlet priors for aspect-topic distributions\\
		$\bm{\gamma_u}, \bm{\gamma_i}$ & Dirichlet priors for aspect distributions\\
		$\bm{\beta_w} $ & Dirichlet priors for topic-word distributions\\
		$\bm{\theta_{u,a}}$ & user's aspect-topic distribution: denoting user's preference on $a$ \\
		$\bm{\psi_{i,a}}$ & item's aspect-topic distribution: denoting  item's features on $a$\\
		$\bm{\lambda_u}, \bm{\lambda_v}$ & aspect distributions of user and item, respectively \\
		$\bm{\phi_w}$  & topic-text word distribution  \\
		\noalign{\smallskip}\hline\noalign{\smallskip}
		$f$    & number of latent factors in ALFM \\
		$\mu_\cdot$ & regularization coefficients \\
		$b_\cdot$ & bias terms, e.g., $b_u, b_i, b_0$ \\
		$w_a$ & weight vector for aspect $a$ \\
		$p_u$, $q_i$ & latent factors of user $u$ and item $i$, respectively \\
		$r_{u,i}$ & rating of user $u$ to item $i$ \\
		$r_{u,i, a}$ & aspect rating  of user $u$ towards item $i$ on aspect $a$\\
		$\rho_{u,i,a}$ & aspect importance of $a$ for $u$ with respect to $i$ \\
		$s_{u,i,a}$   & the degree of item $i$'s attributes matching user $u$'s preference \\
		&  on aspect $a$\\
		\noalign{\smallskip}\Xhline{1.2pt}
	\end{tabular}
	\label{tab:notation}
	\vspace{-0.4cm}
\end{table}

\vspace{-2pt}
\section{Proposed Model} \label{sec:ourmodel}
\subsection{Problem Setting}
Let $\mathcal{D}$ be a collection of reviews of item set $\mathcal{I}$ from a specific category (e.g., restaurant) written by a set of users $\mathcal{U}$, and each review comes with an overall rating $r_{u,i}$ to indicate the overall satisfaction of  user $u$ to item $i$.  The primary goal is to predict the unknown ratings of items that the users have not reviewed yet. A review $d_{u,i}$ is a piece of text which describes opinions of user $u$ on different aspects $a \in \mathcal{A}$ towards item $i$, such as \emph{food} for \emph{restaurants}. In this paper, we only consider the case that all the items are from the same category, i.e., they share the same set of aspects $\mathcal{A}$. Aspects that users care for items are latent and learned from reviews by our proposed topic model, in which each aspect is represented as a distribution of the same set (e.g., $K$) of latent topics. Table~\ref{tab:notation} lists the key notations. Before introducing our method, we would like to first clarify the  concepts of \emph{aspects}, \emph{latent topics}, and \emph{latent factors}. 
\begin{itemize}[align=left,style=nextline,leftmargin=*,labelsep=\parindent,font=\normalfont]
	\item \textbf{Aspect} - it is a high-level semantic concept, which represents the attribute of items that users commented on in reviews,  such as \emph{``food''} for \emph{restaurant} and \emph{``battery"} for \emph{mobile phones}.
	
	\item \textbf{Latent topic  \& latent factor} - in our context, both concepts represent a more fine-grained concept than \emph{``aspect"}. A latent topic or factor can be regarded as a \emph{sub-aspect} of an item. For instance, for the ``food" aspect, a related latent topic could be ``\emph{breakfast}" or ``\emph{Italian cuisine}". We adopt the terminology of \emph{latent topic} in topic models and \emph{latent factor} in matrix factorization.  Accordingly, ``latent topics" are discovered by topic model on reviews, and ``latent factors" are learned by matrix factorization on ratings. 
\end{itemize}

\subsection{Aspect-aware Latent Factor Model}
Based on the observations that (1) different users may care for different aspects of an item and (2) users' preferences may differ from each other for the same aspect, we claim that the overall satisfaction of a user $u$ towards an item $i$ (i.e., the overall rating $r_{u,i}$) depends on $u$'s satisfaction on each aspect $a$ of $i$ (i.e., \emph{aspect rating} $r_{u,i,a}$) and the importance of each aspect (of $i$) to $u$ (i.e., \emph{aspect importance} $\rho_{u,i,a}$). Based on the assumptions, the overall rating $r_{u,i}$ can be predicted as:
\begin{equation}
	\vspace{-2pt}
	\hat{r}_{u,i} = \sum_{a\in\mathcal{A}}  \overbrace{\rho_{u,i,a}}^{\xsub{aspect importance}} \underbrace{r_{u,i,a}}_{\xsub{aspect  rating}}
	\vspace{-2pt}
\end{equation}

\subsubsection{Aspect rating estimation.}
Aspect rating (i.e., $r_{u,i,a}$) reflects the satisfaction of a user $u$ towards an item $i$ on the aspect $a$. To receive a high aspect rating $r_{u,i,a}$, an item should at least possess the characteristics/attributes in which the user is interested in this aspect. Moreover,  the item should satisfy user's expectations on these attributes in this aspect. In other words, the item's attributes on this aspect should be of high quality such that the user likes it.  Take the ``food" aspect as an example, for a user who likes Chinese cuisines, to receive a high rating on the \emph{``food"} aspect from the user, a restaurant should provide Chinese dishes and the dishes should suit the user's taste. Based on user's text reviews, we can learn users' preferences and items' characteristics on each aspect and measure \emph{how the attributes of an item $i$ on aspect ``$a$" suit a user $u$'s requirements on this aspect}, denoted by $s_{u,i,a}$. We compute $s_{u,i,a}$ based on results of the proposed Aspect-aware Topic Model (ATM) (described in Sect.~\ref{sec:matm}), in which user's preferences and item's characteristics on each aspect are modeled as multinomial distributions of latent topics, denoted by $\bm{\theta_{u,a}}$ and $\bm{\psi_{i,a}}$, respectively. $s_{u,i,a} \in [0,1]$ is then computed as :
\begin{equation} \label{eq:jsd}
	\vspace{-2pt}
	s_{u,i,a} = 1-JSD(\bm{\theta_{u,a}}, \bm{\psi_{i,a}})
	\vspace{-1pt}
\end{equation}
where $JSD(\bm{\theta_{u,a}}, \bm{\psi_{i,a}})$ denotes the Jensen–Shannon divergence~\cite{endres2003new} between $\bm{\theta_{u,a}}$ and $\bm{\psi_{i,a}}$. Notice that a large value of $s_{u,i,a}$ does not mean a high rating $r_{u,i,a}$ - an item providing all the features that a user $u$ requires does not mean that it satisfies $u$'s expectations, since the provided ones could be of low quality. For instance, a restaurant provides all the Chinese dishes the user $u$ likes (i.e., high score $s_{u,i,a}$), but these dishes taste bad from $u$'s opinion (i.e., low rating $r_{u,i,a}$). Therefore, we can expect that for this restaurant:  users discuss its Chinese dishes in reviews with negative opinions and thus give low ratings. Instead of analyzing the review sentiments for aspect rating estimation by using external NLP tools (such as~\cite{zhang2014explicit}), we refer to the matrix factorization (MF)~\cite{koren2009matrix} technique.

MF maps users and items into a latent factor space and represents users' preferences and items' features by  $f$-dim latent factor vectors (i.e., $\bm{p_u} \in \mathbb{R}^{f \times 1}$ and $\bm{q_i} \in \mathbb{R}^{f \times 1}$).   The dot product of the user's and item's vectors ($\bm{p_u}^T\bm{q_i}$) characterizes the user's overall interests on the item's characteristics, and is thus used to predict the rating $r_{u,i}$. To extend MF for aspect rating prediction, we introduce a binary matrix  $\bm{W} \in \mathbb{R}^{f \times A}$ to associate the latent factors to different aspects, where $A$ is the number of aspects considered. We call this model  aspect-aware latent factor model (ALFM), in which the weight vector $\bm{w_a}$ in the $a$-th column of $\bm{W}$ indicates which factors are related to the aspect $a$. Thus, $\bm{p_{u,a}} = \bm{w_a} \odot \bm{p_u}$ denotes user's interests in the aspect $a$, where $\odot$ represents element-wise product between vectors. Therefore, $(\bm{p_{u,a}})^T(\bm{q_{i,a}})$ represents the aspect rating of user $u$ to item $i$ on aspect $a$. Finally, we integrate the matching results of aspects  (i.e., $s_{u,i,a}$) into ALFM to estimate the aspect ratings:
\begin{equation}
	\small
	\vspace{-2pt}
	r_{u,i,a} = s_{u,i,a} \cdot (\bm{w_a} \odot \bm{p_u})^T(\bm{w_a} \odot \bm{q_i})
	\vspace{-1pt}
\end{equation}
As a high aspect rating $r_{u,i,a}$ requires  large values of both $s_{u,i,a}$ and $(\bm{w_a} \odot \bm{p_u})^T(\bm{w_a} \odot \bm{q_i})$, it is expected that the results learned from reviews could guide the learning of latent factors.

\subsubsection{Aspect importance estimation.}
We rely on user reviews to estimate $\rho_{u,i,a}$, as users often discuss their interest topics of aspects in reviews, such as different \emph{cuisines} in the \emph{food} aspect. In general, the more a user comments on an aspect in reviews, the more important this aspect is (to this user). Thus, we estimate the importance of an aspect according to the possibility of a user writing review comments on this aspect. When writing a review,  some users tend to write comments from the aspects according to their own preferences, while others like commenting on the most notable features of the targeted item. Based on this consideration, we introduce (1) $\pi_u$ to denote the probability of user $u$ commenting an item based on his own preference and (2) $\lambda_{u,a}$ ($\sum_{a\in\mathcal{A}}\lambda_{u,a}=1$) to denote the probability of user $u$ commenting on the aspect $a$ based on his own preference. Accordingly, $(1-\pi_{u})$ denotes the probability of the user commenting from the item $i$'s characteristics ($\sum_{a\in\mathcal{A}}\lambda_{i,a}=1$), and $\lambda_{i,a}$ is the probability of user $u$ commenting item $i$ from the item's characteristics on the aspect $a$. Thus, the probability of a user $u$ commenting an item $i$ on an aspect $a$ (i.e., $\rho_{u,i,a}$) is:
\begin{equation} \label{eq:rho}
	\small
	\vspace{-2pt}
	\rho_{u,i,a} = \pi_{u}\lambda_{u,a} + (1-\pi_{u})\lambda_{i,a}
	\vspace{-1pt}
\end{equation}
$\lambda_{u,a}$, $\lambda_{i,a}$, and $\pi_u$ are estimated by ATM, which simulates the process of a user writing a review, as detailed in the next subsection.


\subsection{Aspect-aware Topic Model} \label{sec:matm}

Given a corpus $\mathcal{D}$, which contains reviews of users towards items $\{d_{u,i}|d_{u,i} \in \mathcal{D}, u  \in \mathcal{U}, i \in \mathcal{I}\}$, we assume that  a set of latent topics (i.e., $K$ topics) covers all the topics that users discuss in the reviews. $\bm{\lambda_u}$ is a probability distribution of aspects in user $u$'s preferences, in which each value  $\lambda_{u,a}$ denotes the relative importance of an aspect $a$ to the user $u$. Similarly, $\bm{\lambda_i}$ is the probability distribution of aspects in item $i$'s characteristics, in which each value  $\lambda_{i,a}$ denotes the importance of an aspect $a$ to the item $i$.  As the $K$ latent topics cover all the topics discussed in reviews,  an aspect will only  relate to some of the latent topics closely. For example, topic ``\emph{breakfast}" is closely related to aspect  ``food", while it is not related to aspects like ``\emph{service}" or ``\emph{price}". The relation between aspects and topics is also represented by a probabilistic distribution, i.e.,   $\bm{\theta_{u,a}}$  for users and $\bm{\psi_{i,a}}$ for items. More detailedly, the interests of a user $u$ in a specific aspect $a$ is represented by $\bm{\theta_{u,a}}$, which is a multinomial distribution of the latent topics; the characteristics of an item $i$ in a specific aspect $a$ is represented by $\bm{\psi_{i,a}}$, which is also a multinomial distribution of the same set of latent topics.
$\bm{\theta_{u,a}}$ is determined  based on all the reviews $\{d_{u,i} | i \in \mathcal{I}\}$ of user $u$ writing for items. $\bm{\psi_{i,a}}$ is learned from all the reviews $\{d_{u,i} | u \in \mathcal{U}\}$ of $i$ written by users.  A latent topic is a multinomial distribution of text words in reviews. Based on these assumptions, we propose an aspect-aware topic model ATM to estimate the parameters $\{\bm{\lambda_i}$, $\bm{\lambda_i}$, $\bm{\theta_{u,a}}$, $\bm{\psi_{i,a}}$, $\pi_u\}$ by simulating the generation of the corpus $\mathcal{D}$.

\begin{figure}
	\centering
	\small
	\includegraphics[width = 7cm]{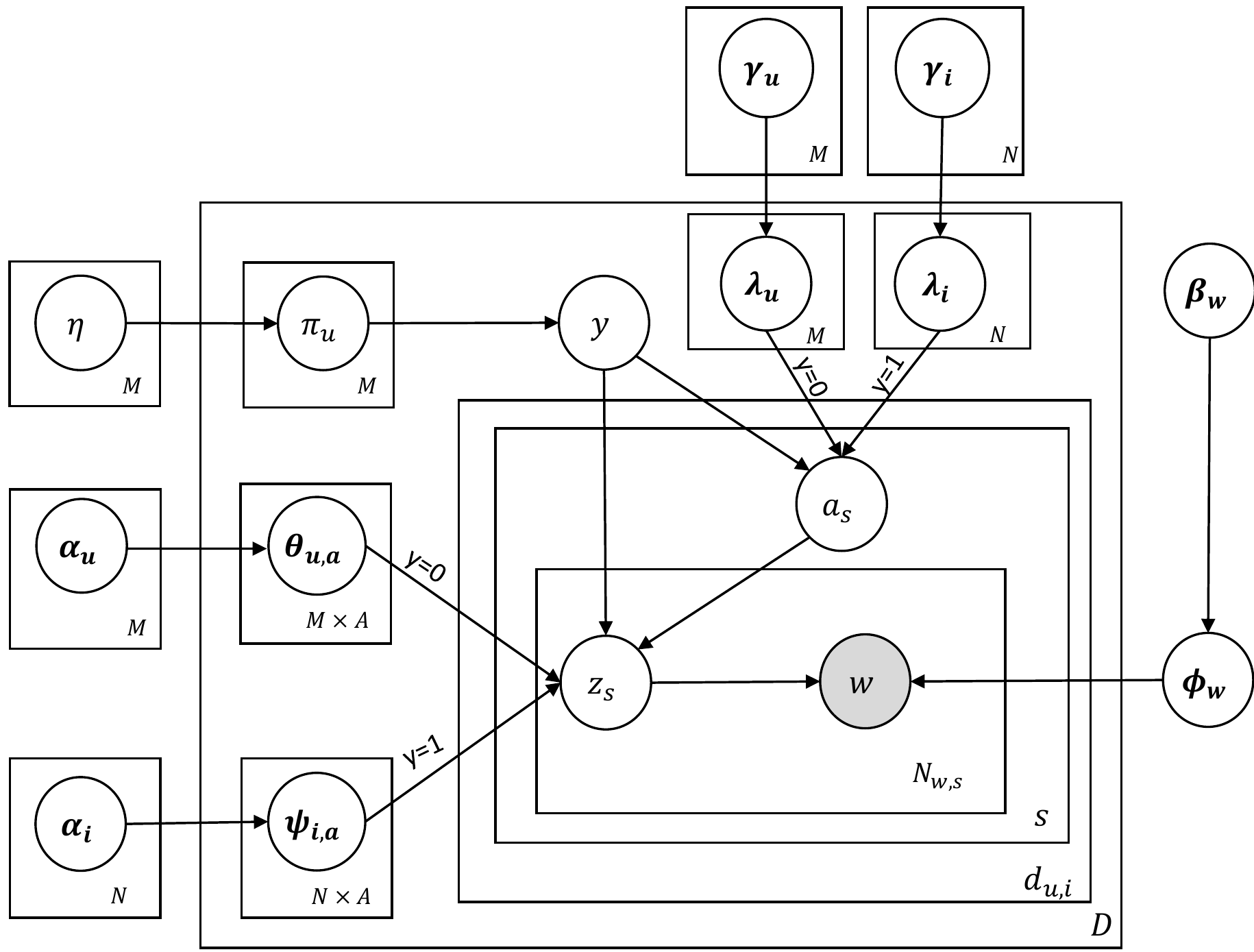}
	\caption{The graphical representation of the ATM model. }
	\label{fig:matm}
	\vspace{-0.4cm}
\end{figure}
The graphical representation of ATM is shown in Fig.~\ref{fig:matm}. In the figure, the shaded circles indicate observed variables, while the unshaded ones represent the latent variables.  ATM mimics the processing of writing a review sentence by sentence. A sentence usually discusses the same topic $z$, which could be from user's preferences or from item's characteristics.  To decide the topic $z_s$ for a sentence $s$, our model introduces  an indicator variable  $y \in \{0,1\}$ based on a Bernoulli distribution, which is parameterized by $\pi_u$. Specifically,  when $y=0$, the sentence is generated from user's preference; otherwise, it is generated according to item $i$'s characteristics.   $\pi_u$ is user-dependent, indicating the tendency to comment from $u$'s personal preferences or from the item $i$'s characteristics is determined by $u$'s personality. The generation process of ATM is shown in  Algorithm~\ref{alg:geneproc}. Let $a_s$ denote the aspect assigned to a sentence $s$.  If $y=0$, $a_s$ is drawn from $\lambda_u$ and $z_s$ is then generated from $u$'s preferences on aspect $a_s$: $\bm{\theta_{u,a_s}}$; otherwise, if $y=1$,  $a_s$ is drawn from $\lambda_i$ and $z_s$ is then generated from $i$'s characteristics on aspect $a_s$: $\bm{\psi_{i,a_s}}$. Then all the words $w$ in sentence $s$ is generated from $z_s$ according to the word distribution: $\bm{\phi_{z_s, w}}$.

\begin{algorithm}
	\footnotesize
	\For{each topic $k = 1, ..., K$ }
	{
		Draw $\bm{\phi_{k,w}} \sim Dir(\cdot|\bm{\beta_w})$\;
	}
	\For{each user $u \in \mathcal{U}$}
	{
		Draw $\bm{\lambda_u} \sim Dir(\cdot|\bm{\gamma_u})$\;
	}
	\For{each item $i \in \mathcal{I}$}
	{
		Draw $\bm{\lambda_i} \sim Dir(\cdot|\bm{\gamma_i})$\;
	}
	\For{each user $u \in \mathcal{U}$, each aspect $a \in \mathcal{A}$}
	{
		Draw $\bm{\theta_{u,a}} \sim Dir(\cdot|\bm{\alpha_u})$\;
	}
	\For{each item $i \in \mathcal{I}$, each aspect $a \in \mathcal{A}$}
	{
		Draw $\bm{\psi_{i,a}} \sim Dir(\cdot|\bm{\alpha_i})$\;
	}
	\For{each review $d_{u,i}, u \in \mathcal{U}, i \in \mathcal{I}$ }
	{
		\For{each sentence $s \in d_{u,i}$ }
		{
			Draw $y \sim Bernoulli(\cdot|\pi_u)$\;
			\If{$y_s == 0$}{Draw $a_s \sim Multi(\bm{\lambda_u})$ and then draw $z_s \sim Multi(\bm{\theta_{u,a_s}})$ \;}
			\If{$y_s == 1$}{Draw $a_s \sim Multi(\bm{\lambda_i})$ and then draw $z_s \sim Multi(\bm{\psi_{i,a_s}})$\;}
			
			\For{each word $w \in s$ }
			{
				Draw $w \sim Multi(\phi_{z_s,w})$
			}
		}
	}
	\caption{Generation Process of ATM}
	\vspace{-0.1cm}
	\label{alg:geneproc}
\end{algorithm}

In ATM, $\bm{\alpha_u}$, $\bm{\alpha_i}$, $\bm{\gamma_u}$, $\bm{\gamma_i}$, $\bm{\beta}$, and $\eta$ are pre-defined hyper-parameters and set to be symmetric for simplicity. Parameters need to be estimated including $\bm{\lambda_i}$, $\bm{\lambda_i}$, $\bm{\theta_{u,a}}$, $\bm{\psi_{i,a}}$, and $\pi_u$. Different approximate inference methods have been developed for parameter estimation in topic models, such as variation inference~\cite{blei2003latent} and collapsed Gibbs sampling~\cite{griffiths2004}. We apply collapsed Gibbs sampling to infer the parameters, since it has been successfully applied in many large scale applications of topic models~\cite{cheng2016effective,cheng2017sigir}.  Due to the space limitation, we omit the detailed inference steps in this paper.

\subsection{Model Inference} With the results of ATM, $\rho_{u,i,a}$ and $s_{u,i,a}$ can be computed using Eq.~\ref{eq:rho} and ~\ref{eq:jsd}, respectively. With the consideration of bias terms (i.e., $b_u, b_i, b_0$) in ALFM, the overall rating can be estimated as\footnote{In our experiments, we tried to normalize $\rho_{u,i,a}$ or $\rho_{u,i,a} \cdot s_{u,i,a}$ in Eq.~\ref{eq:re2}, but no improvement has been observed.},
\begin{equation} \label{eq:re2}
	\hat{r}_{u,i} = \sum_{a\in\mathcal{A}}(\rho_{u,i,a}\cdot s_{u,i,a} \cdot (\bm{w_a} \odot \bm{p_u})^T(\bm{w_a} \odot \bm{q_i}) )  + b_u + b_i + b_0
\end{equation}
where $b_0$ is the average rating,  $b_u$ and $b_i$ are user and item biases, respectively. The estimation of parameters is to minimize the rating prediction error in the training dataset. The optimization objective function is
\vspace{-1pt}
\begin{equation} \label{eq:ojf}
	\begin{split}
		\vspace{-2pt}
		\underset{p*,q*}{\text{min}} \frac{1}{2}\sum_{u,i} (r_{u,i} & -\hat{r}_{u,i})^2 + \frac{\mu_u}{2} ||\bm{p_u}||_2^2 + \frac{\mu_i}{2} ||\bm{q_i}||_2^2 \\
		& + \mu_w \sum_a||\bm{w_a}||_1 + \frac{\mu_b}{2} (||b_u||_2^2 + ||b_i||_2^2);
		\vspace{-2pt}
	\end{split}
\end{equation}
where $||\cdot||_2$ denotes the $\ell_2$ norm for preventing model overfitting, and $||\cdot||_1$ denotes the $\ell_1$ norm.  $\mu_u, \mu_i, \mu_w$, and $\mu_b$ are regularization parameters, which are tunable hyper-parameters. In practice, we relax the binary requirement of $\bm{w_a}$ by using $\ell_l$ norm. It is well known that  $\ell_l$ regularization yields sparse solution of the weights~\cite{mairal2010online}. The $\ell_2$ regularization of $\bm{p_u}$ and $\bm{q_i}$ prevents them to have arbitrarily large values, which would lead to arbitrarily small values of $\bm{w_a}$.

\textbf{Optimization.} We use the stochastic gradient descent (SGD) algorithm to learn the parameters by optimizing the objective function in Eq.~\ref{eq:ojf}. In each step of SGD, the localized optimization is performed on a rating $r_{u,i}$.  Let $L$ denote the loss, and the gradients of parameters are given as follows:
\vspace{-0.1cm}
\begin{align*}\label{eq:user}
	\small
	\vspace{-2pt}
	\frac{\partial L}{\partial p_u}&=\sum_{i=1}^{N}(\sum_{a}\rho_{u,i,a}s_{u,i,a}w_a^2)(\hat{r}_{u,i}-r_{u,i})q_i + \mu_u p_u \\
	\frac{\partial L}{\partial q_i}&=\sum_{u=1}^{M}(\sum_{a}\rho_{u,i,a}s_{u,i,a}w_a^2)(\hat{r}_{u,i}-r_{u,i})p_u + \mu_i q_i \\
	\frac{\partial L}{\partial w_a}&=\sum_{u=1}^{M}\sum_{i=1}^{N}\rho_{u,i,a}s_{u,i,a}(\hat{r}_{u,i}-r_{u,i})p_uq_iw_a + \frac{\mu_w w_a} {\sqrt{(w_a^2+\epsilon)}}
	\vspace{-2pt}
\end{align*}
Here, we omit the gradients of $b_u$ and $b_i$, as they are the same as in the standard biased MF~\cite{koren2009matrix}.  $M$ and $N$ are the total number of users and items in the dataset.  Notice that in the deriving of the gradient for $w_a$, we use $\sqrt{w_a^2+\epsilon}$ in place of $||w_a||_1$, because $\ell_1$ norm is not differentiable at 0. $\epsilon$ can be regarded as a ``smoothing parameter" and is set to $10^{-6}$ in our implementation.  


\section{Experimental Study} \label{sec:expconfig}
To validate the assumptions when designing the model and evaluate our proposed model, we conducted comprehensive experimental studies to answer the following questions:
\begin{itemize} [align=left,style=nextline,leftmargin=*,labelsep=\parindent,font=\normalfont]
	\item \textbf{RQ1:} How do the important parameters (e.g., the number of latent topics and latent factors) affect the performance of our model? More importantly, is the setting $f=K$ optimal, which is a default assumption for many previous models? (Sect.~\ref{sec:modelanalysis})
	\item \textbf{RQ2:} Can our ALFM model outperform the state-of-the-art recommendation methods, which consider both ratings and reviews, on rating prediction? (Sect.~\ref{sec:comp})
	\item \textbf{RQ3:} Compared to other methods which also use textual reviews and ratings, how does our ALFM model perform on the cold-start setting when users have only few ratings? (Sect.~\ref{sec:coldstart})
	\item \textbf{RQ4:} Can our model explicitly interpret the reasons for a high or low rating? (Sect.~\ref{sec:interpret})
\end{itemize}

\begin{table}[]
	\small
	\centering
	\caption{Statistics of the evaluation datasets}
	\vspace{-0.2cm}
	\begin{tabular}{ccccc} \toprule
		Datasets	&	\#users	&	\#items	&	\#ratings	&	Sparsity	\\ \hline
		Instant Video	&	4,902	&	1,683	&	36,486	&	0.9956	\\
		Automotive	&	2,788	&	1,835	&	20,218	&	0.9960	\\
		Baby	&	17,177	&	7,047	&	158,311	&	0.9987	\\
		Beauty	&	19,766	&	12,100	&	196,325	&	0.9992	\\
		Cell Phones 	&	24,650	&	10,420	&	189,255	&	0.9993	\\
		Clothing	&	34,447	&	23,026	&	277,324	&	0.9997	\\
		Digital Music	&	5,426	&	3,568	&	64,475	&	0.9967	\\
		Grocery 	&	13,979	&	8,711	&	149,434	&	0.9988	\\
		Health 	&	34,850	&	18,533	&	342,262	&	0.9995	\\
		Home \& Kitchen	&	58,901	&	28,231	&	544,239	&	0.9997	\\
		Musical Instruments	&	1,397	&	900	&	10,216	&	0.9919	\\
		Office Products	&	4,798	&	2,419	&	52,673	&	0.9955	\\
		Patio	&	1,672	&	962	&	13,077	&	0.9919	\\
		Pet Supplies	&	18,070	&	8,508	&	155,692	&	0.9990	\\
		Sports \& Outdoors	&	31,176	&	18,355	&	293,306	&	0.9995	\\
		Tools \& Home 	&	15,438	&	10,214	&	133,414	&	0.9992	\\
		Toys \& Games	&	17,692	&	11,924	&	166,180	&	0.9992	\\
		Video Games	&	22,348	&	10,672	&	228,164	&	0.9990	\\
		Yelp 2017	&	169,257	&	63,300		&	1,659,678	&	0.9998	\\ \hline
	\end{tabular}
	\label{tab:dataset}
	\vspace{-0.2cm}
\end{table}

\subsection{Dataset Description} \label{sec:dataset}
We conducted experiments on two publicly accessible datasets that provide user review and rating information. The first dataset is
Amazon Product Review dataset collected by~\cite{mcauley2013hidden}\footnote{http://jmcauley.ucsd.edu/data/amazon/}, which contains product reviews and metadata from Amazon. This dataset has been widely used for rating prediction with reviews and ratings in previous studies~\cite{mcauley2013hidden,ling2014ratings,tan2016rating,catherine2017transnets}. The dataset is organized into 24 product categories. In this paper, we used 18 categories (See Table~\ref{tab:dataset}) and focus on the 5-core version, with at least 5 reviews for each user or item.  The other dataset is from Yelp Dataset Challenge 2017\footnote{http://www.yelp.com/dataset\_challenge/}, which includes reviews of local business in 12 metropolitan areas across 4 countries. For the Yelp 2017 dataset, we also processed it to keep users and items with at least 5 reviews.  From each review in these datasets, we extract the corresponding ``userID", ``itemID", a rating score (from 1 to 5 rating stars), and a textual review for experiments. Notice that for all the datasets, we checked and removed the duplicates, and then filtered again to keep them as 5-core. Besides, we removed the infrequent terms in the reviews for each dataset.\footnote{The thresholds of infrequent terms varied across different datasets. For example, for the ``Yelp 2017" dataset, which is relatively large, a term that appears less than 10 times in reviews is defined as an infrequent term; and the thresholds are smaller for relatively small datasets (e.g., the threshold is 5 for the ``Music Instruments" dataset).)}  Some statistics of the datasets are shown in Table~\ref{tab:dataset}. 

\subsection{Experimental Settings}
For each dataset, we randomly split it into training, validation, and testing set with ratio 80:10:10 for each user as in~\cite{mcauley2013hidden,ling2014ratings,catherine2017transnets}. Because we take the 5-core dataset where each user has at least 5 interactions, we have at least 3 interactions per user for training, and at least 1 interaction per user for validation and testing. Note that we only used the review information in the training set, because the reviews in the validation or testing set are unavailable during the prediction process in real scenarios. The number of aspect is set to 5 in experiments.\footnote{We tuned the number of aspects from 1 to 8 for all the datasets, and found that the performance does not change much unless setting the aspect number to 1 or 2. }

\textbf{Baselines:} We compare the proposed \textbf{ALFM} model with the following baselines. It is worth noting that these methods are tuned on the validation dataset to obtain their optimal hyper-parameter settings for fair comparisons.

\begin{itemize} [align=left,style=nextline,leftmargin=*,labelsep=\parindent,font=\normalfont]
	\item \textbf{BMF~\cite{koren2009matrix}.} It is a standard MF method with the consideration of bias terms (i.e., user biases and item biases). This method only leverages ratings when modeling users' and items' latent factors.  It is typically a strong baseline model in collaborative filtering~\cite{koren2009matrix,ling2014ratings}. 
	
	\item \textbf{HFT~\cite{mcauley2013hidden}.} It models ratings with MF and review text with latent topic model (e.g., LDA~\cite{blei2003latent}). We use it as a representative of the methods which use an exponential transformation function  to link the latent topics with latent factors, such as TopicMF~\cite{bao2014topicmf}.  The topic distribution can be modeled on either users or items. We use the topic distribution based on items, since it achieves better results. Note that in experiments, we add bias terms into HFT, which can achieve better performance.
	
	\item \textbf{CTR~\cite{wang2011collaborative}.} This method also utilizes both review and rating information. It uses a topic model to learn the topic distribution of items, which is then used as the latent factors of items in MF with an addition of a latent variable. 
	
	\item \textbf{RMR~\cite{ling2014ratings}.} This method also uses both ratings and reviews. Different from HFT and CTR, which use MF to model rating, it uses a mixture of Gaussian to model the ratings.  
	
	\item \textbf{RBLT~\cite{tan2016rating}.} This method is the most recent method, which also uses MF to model ratings and LDA to model review texts. Instead of using an exponential transformation function to link the latent topics and latent factors (as in HFT~\cite{mcauley2013hidden}), this method linearly combines the latent factors and latent topics  to represent users and items, with the assumption that the dimensions of topics and latent factors are equal and in the same latent space. The same strategy is also adopted by ITLFM~\cite{zhang2016integrating}. Here, we use RBLT as a representative method for this strategy.
	
	\item \textbf{TransNet~\cite{catherine2017transnets}.} This method adopts neural network frameworks for rating prediction. In this model, the reviews of users and items are used as input to learn the latent representations of users and items. More descriptions about this method could be found in Section~\ref{sec:relwork}. We used the codes published by the authors in our experiments and tuned the parameters as described in~\cite{catherine2017transnets}.
	
\end{itemize}


The standard root-mean-square error (\textbf{RMSE}) is adopted in evaluation. A smaller RMSE value indicates better performance.


\begin{table*}[]
	\small
	\centering
	\caption{Comparisons of adopted methods in terms of RMSE with $f=K=5$.}
	\vspace{-0.2cm}
	\begin{tabular}{|c|C{0.6cm}C{0.6cm}C{0.6cm}C{0.6cm}C{0.6cm}C{0.9cm}C{0.75cm}|C{0.87cm}C{0.89cm}C{0.87cm}C{0.89cm}C{0.95cm}C{0.85cm}|} \hline
		\multirow{2}{*}{Datasets}	&	BMF	&	HFT	&	CTR	&	RMR	&	RBLT	&	TransNet	&	ALFM	&	\multicolumn{6}{c|}{Improvement(\%)}											\\  \cline{9-14}
		&	(a)	&	(b)	&	(c)	&	(d)	&	(e)	&	(f)	&	(g)	&	g vs. a 	&	g  vs. b 	&	g  vs. c 	&	g vs. d 	&	g  vs. e 	&	g  vs. f \\  \hline
		Instant Video	&	1.162	&	0.999	&	1.014	&	1.039	&	0.978	&	0.996	&	\textbf{0.967}	&	16.79	&	3.19	&	4.63*	&	6.94*	&	1.12**	&	2.88	\\
		Automotive	&	1.032	&	0.968	&	1.016	&	0.997	&	0.924	&	0.918	&	\textbf{0.885}	&	14.26*	&	8.58**	&	12.86*	&	11.19*	&	4.24**	&	3.56*	\\
		Baby	&	1.359	&	1.112	&	1.144	&	1.178	&	1.122	&	1.110	&	\textbf{1.076}	&	20.83**	&	3.24	&	5.98*	&	8.66**	&	4.11	&	3.05*	\\
		Beauty	&	1.342	&	1.132	&	1.171	&	1.190	&	1.117	&	1.123	&	\textbf{1.082}	&	19.39**	&	4.47	&	7.65**	&	9.12**	&	3.18**	&	3.65**	\\
		Phones 	&	1.432	&	1.216	&	1.271	&	1.289	&	1.220	&	1.207	&	\textbf{1.167}	&	18.47**	&	3.98*	&	8.18*	&	9.4**	&	4.33	&	3.27**	\\
		Clothing	&	1.073	&	1.103	&	1.142	&	1.145	&	1.073	&	1.064	&	\textbf{1.032}	&	3.8**	&	6.47**	&	9.65	&	9.9**	&	3.86**	&	2.96*	\\
		Digital Music	&	1.093	&	\textbf{0.918}	&	0.921	&	0.960	&	\textbf{0.918}	&	1.061	&	0.920	&	15.82	&	-0.15	&	0.13*	&	4.49**	&	-0.15**	&	4.13**	\\
		Grocery 	&	1.192	&	1.016	&	1.045	&	1.061	&	1.012	&	1.022	&	\textbf{0.982}	&	17.66**	&	3.36**	&	6.07	&	7.46**	&	3.01**	&	3.94*	\\
		Health 	&	1.263	&	1.073	&	1.105	&	1.135	&	1.070	&	1.114	&	\textbf{1.042}	&	17.48*	&	2.83	&	5.65*	&	8.20	&	2.56**	&	6.46**	\\
		Home \& Kitchen	&	1.297	&	1.083	&	1.123	&	1.149	&	1.086	&	1.123	&	\textbf{1.049}	&	19.16**	&	3.15**	&	6.62	&	8.7**	&	3.41**	&	6.61**	\\
		Musical Instruments	&	1.004	&	0.972	&	0.979	&	0.983	&	0.946	&	0.901	&	\textbf{0.893}	&	11.08	&	8.17**	&	8.83**	&	9.2**	&	5.61	&	0.95	\\
		Office Products	&	1.025	&	0.879	&	0.898	&	0.934	&	0.872	&	0.898	&	\textbf{0.848}	&	17.29**	&	3.55**	&	5.61*	&	9.26**	&	2.77**	&	5.67**	\\
		Patio	&	1.180	&	1.041	&	1.062	&	1.077	&	1.032	&	1.046	&	\textbf{1.001}	&	15.19**	&	3.84*	&	5.7*	&	7.07*	&	2.96	&	4.33**	\\
		Pet Supplies	&	1.367	&	1.137	&	1.177	&	1.200	&	1.139	&	1.149	&	\textbf{1.099}	&	19.64**	&	3.41*	&	6.67*	&	8.41	&	3.54**	&	4.38**	\\
		Sports \& Outdoors	&	1.130	&	0.970	&	0.998	&	1.019	&	0.964	&	0.990	&	\textbf{0.933}	&	17.42**	&	3.8*	&	6.47	&	8.4*	&	3.2**	&	5.77**	\\
		Tools \& Home	&	1.168	&	1.013	&	1.047	&	1.090	&	1.011	&	1.041	&	\textbf{0.974}	&	16.63**	&	3.90	&	6.98	&	10.68**	&	3.7**	&	6.51**	\\
		Toys \& Games	&	1.072	&	0.926	&	0.948	&	0.974	&	0.923	&	0.951	&	\textbf{0.902}	&	15.81**	&	2.59*	&	4.82**	&	7.39**	&	2.3**	&	5.11*	\\
		Video Games	&	1.321	&	1.096	&	1.115	&	1.150	&	1.094	&	1.123	&	\textbf{1.070}	&	19.02*	&	2.43	&	4.03**	&	6.97*	&	2.24**	&	4.77*	\\
		Yelp 2017	&	1.415	&	1.174	&	1.233	& 1.266	&	1.202	&	1.190	&	\textbf{1.155}	&	18.35*	&	1.60**	&	6.33**	&	8.74*	&	3.88**	&	2.92*	\\  \hline
		Average	&	1.207	&	1.044	&	1.074	&	1.097	&	1.037	&	1.049	&	\textbf{1.004}	&	14.56**	&	2.84*	&	7.16**	&	8.31*	&	3.37**	&	4.26**	\\  \hline
	\end{tabular}
	\begin{tablenotes}
		\footnotesize
		\item The improvements with * are significant with $p-value < 0.05$, and the improvements with ** are significant with $p-value < 0.01$ with a two-tailed paired t-test.
	\end{tablenotes}
	\label{tab:comp}
	\vspace{-0.4cm}
\end{table*}

\begin{figure}[]
	\centering
	\hspace{-0.3cm}
	\includegraphics[height = 4.2cm]{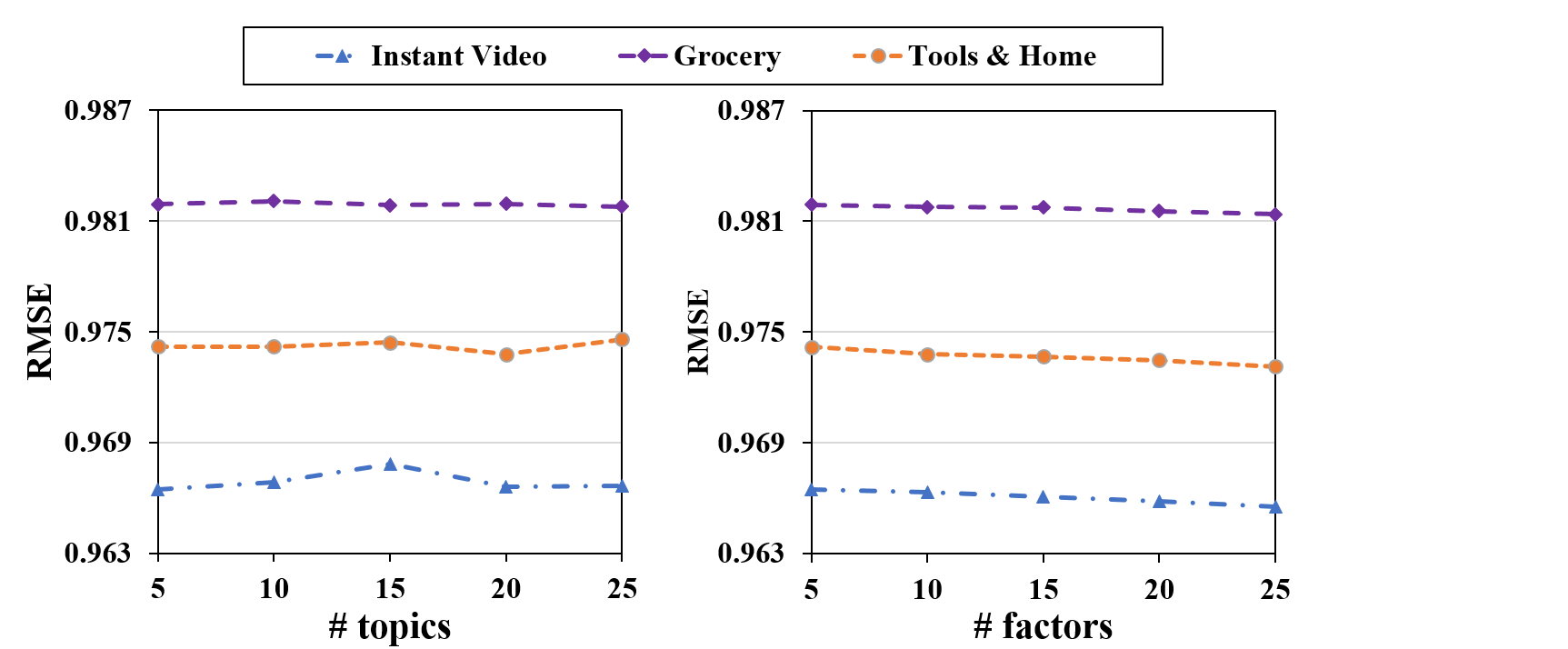}
	\vspace{-0.3cm}
	\caption{Effects of factors and topics in our model.}
	\label{fig:effects}
\end{figure}

\begin{figure}[]
	\centering
	\subfloat[Clothing]{
		\includegraphics[height = 3.3cm]{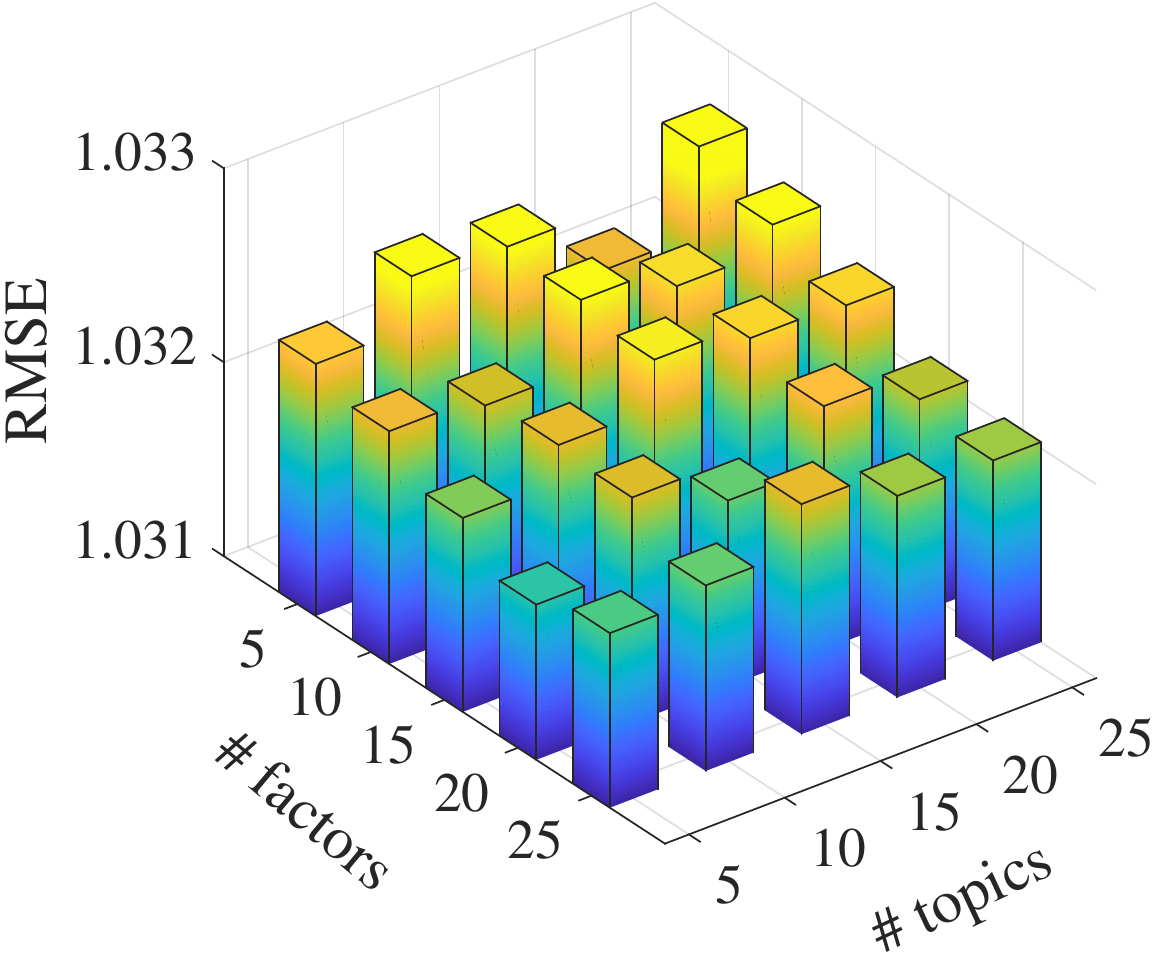}
	}
	\subfloat[Yelp 2017]{
		\includegraphics[height = 3.3cm]{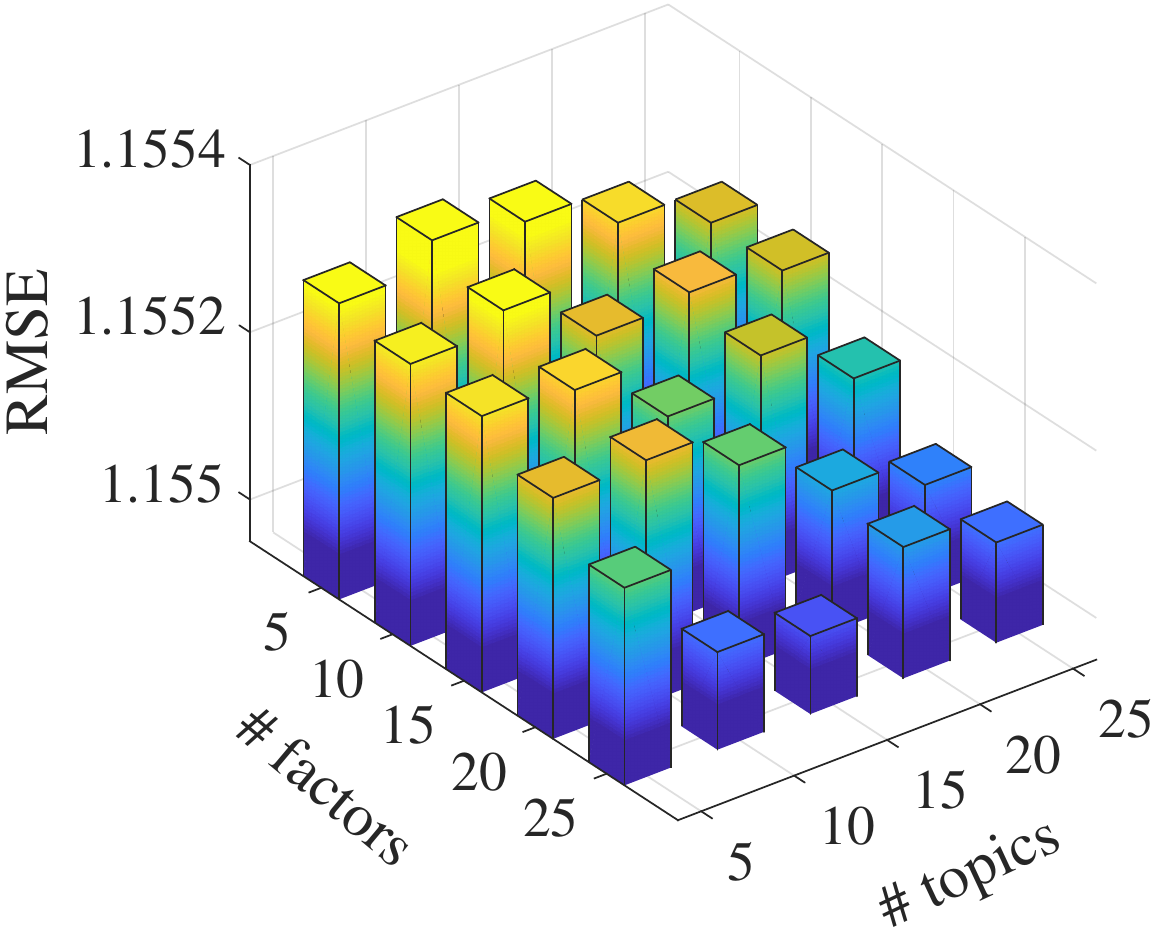}
	}	
	\vspace{-0.3cm}	
	\caption{Effects of \#factors v.s. \#topics. }
	\label{fig:factorvstopic}
\end{figure}

\vspace{-0.1cm}

\subsection{Effect of Important Parameters (RQ1)} \label{sec:modelanalysis}
In this subsection, we analyze the influence of \emph{the number of  latent factors} and \emph{the number of latent topics} on the final performance of ALFM.
As we know, in MF, more  latent factors will lead to better performance unless overfitting occurs~\cite{he2016fast,koren2009matrix}; while the optimal number of latent topics in topic models (e.g., LDA) is dependent on the datasets~\cite{blei2012probabilistic,arun2010finding}. Accordingly, the optimal number of latent topics in topic model and the optimal number of latent factors in MF should be tuned separately. However, in the previous latent factor models (e.g., HFT, TopicMF~\cite{bao2014topicmf}, RMR, CTR, and RBLT), the number of factors (i.e., \#factors) and the number of topics (i.e., \#topics) are assumed to be the same, and thus cannot be optimized separately. Since our model does not have such constraint, we studied the effects of \#factors and \#topics individually. Fig.~\ref{fig:effects} show the performance variations with the change of \#factors and \#topics  by setting the other one to 5. We only visualize the performance variations of three datasets, due to the space limitation and the similar performance variation behaviors of other datasets. 
From the figure, we can see that with the increase of \#factors, RMSE consistently decreases although the degree of decline is small. Notice that in our model, the rating prediction still relies on MF technique (Eq.~\ref{eq:re2}). Therefore, the increase of  \#factors could lead to better representation capability and thus more accurate prediction. In contrast, the optimal number of latent topics is different from dataset to dataset.

To better visualize the impact of \#factors and \#topics, we also present 3D figures by varying the number of factors and topics in $\{5, 10, 15, 20, 25\}$, as shown in Fig.~\ref{fig:factorvstopic}.  In this figure, we use the performance of three datasets as illustration. From the figure, we can see that the optimal numbers of topics and latent factors are varied across different datasets. In general, more latent factors usually lead to better performance, while the optimal number of latent topics is dependent on the reviews of different datasets. This also reveals that setting \#factors  and \#topics to be the same may not be optimal.

\subsection{Model Comparison (RQ2)} \label{sec:comp}
We show the performance comparisons of our ALFM with all the baseline methods in Table~\ref{tab:comp}, where the best prediction result on each dataset is in bold. For fair comparison, we set the number of latent factors ($f$) and the number of latent topics ($K$) to be the same as $f=K=5$. Notice that our model could obtain better performance when setting $f$ and $K$ differently. Still, ALFM achieves the best results on 18 out of the 19 datasets. Compared with BMF, which only uses ratings, we achieve much better prediction performance (16.49\% relative improvement on average). More importantly, our model outperforms CTR and RMR with large margins - 6.28\% and 8.18\% relative improvements on average, respectively. Compared to the recently proposed RBLT and TransNet, ALFM can still achieve 3.37\% and 4.26\% relative improvement on average respectively with significance testing.   It is worth mentioning that HFT achieves better performance than RMR and comparable performance with recent RBLT, because we added bias terms to the original HFT in~\cite{mcauley2013hidden}. TransNet applies neural networks, which has exhibited strong capabilities on representation learning, in reviews to learn users' preferences and items' characteristics for rating prediction. However, it may suffer from (1) noisy information in reviews, which would deteriorate the performance; and (2) errors introduced when generating fake reviews for rating prediction, which  will also cause bias in the final performance.
Compared to those baselines, the advantage of ALFM  is that it models users' preferences on different aspects; and more importantly, it captures a user's specific attention on each aspect of a targeted item. The substantial improvement of ALFM over those baselines demonstrates the benefits of modeling users' specific preferences on each aspect of different items.

\begin{figure*}[]
	\centering
	\hspace{-0.3cm}
	\subfloat[Clothing]{
		\includegraphics[height = 4cm]{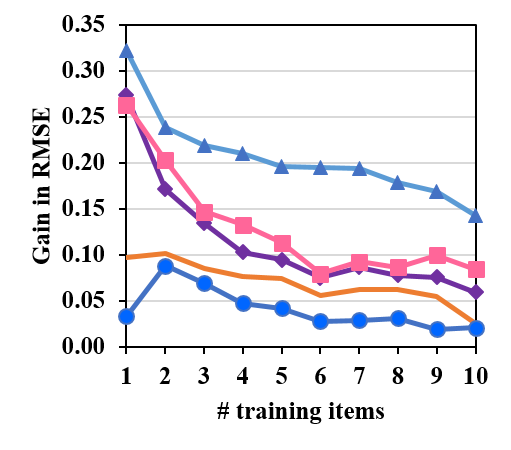}
	}
	\subfloat[Yelp 2017]{
		\includegraphics[height = 4cm]{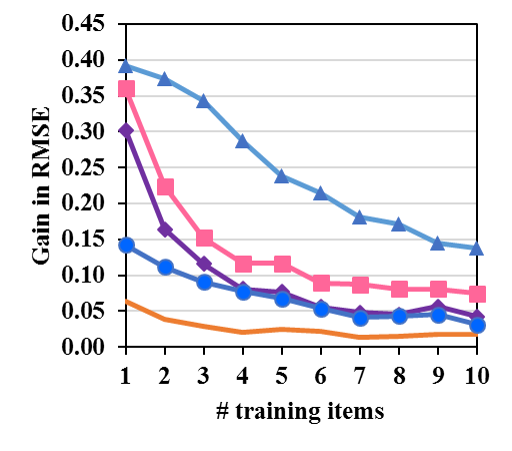}          
	}	
	\subfloat[All Datasets]{
		\includegraphics[height = 4cm]{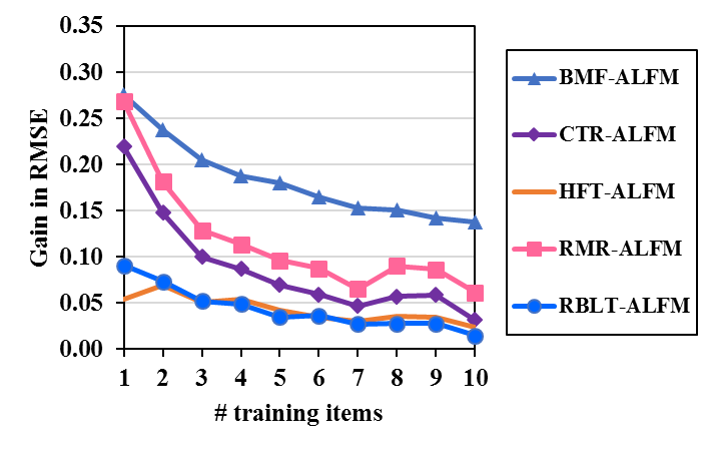}
	}	
	\vspace{-0.3cm}
	\caption{Gain in RMSE for user with limited training data on two individual datasets and overall 19 datasets.}
	\label{fig:coldstart}
	
\end{figure*}
\vspace{-0.2cm}

\subsection{Cold-Start Setting (RQ3)} \label{sec:coldstart}
As shown in Table~\ref{tab:dataset}, the datasets are usually very sparse in practical systems. It is inherently difficult to provide satisfactory recommendation based on limited ratings. In the matrix factorization model, given a few ratings, the penalty function tends to push $q_u$ and $p_i$ towards zero. As a result, such users and items are modeled only with the bias terms~\cite{ling2014ratings}. Therefore, matrix factorization easily suffers from the cold-start problem. By integrating reviews in users' and items' latent factor learning, our model could alleviate the problem of cold-start to a great extent, since reviews contain rich information about user preferences and item features.

To demonstrate the capability of our model in dealing with users with very limited ratings, we randomly split the datasets into training, validation, and testing sets in ratio 80:10:10 based on the number of ratings in each set. In this setting, it is not guaranteed  that a user has at least 3 ratings in the training set. It is possible that a user has no rating in the training set. For the users without any ratings in the training set, we also removed them in the testing set.  Then we evaluate the performance of users who have the number of ratings from 1 to 10 in the training set.
In Fig.~\ref{fig:coldstart}, we show the \textbf{Gain in RMSE} ($y$-axis) grouped by the number of ratings ($x$-axis) of users in the training set.  The value of \textbf{Gain in RMSE} is equal to the average RMSE of baselines \emph{minus} that of
our  model (e.g., ``BMF-TALFM"). A positive value indicates that our model achieves better prediction. As we can see, our ALFM model substantially improves the prediction accuracy compared with the BMF model. More importantly, our model also outperforms all the other baselines which also utilize reviews. This demonstrates that our model is more effective in exploiting reviews and ratings,  because it learns user's preferences and item's features in different aspects and is capable of estimating the aspect weights based on the targeted user's preferences and targeted item's features. 

\begin{table}[]
	\small
	\centering
	\caption{Top ten words of each aspect for a user (index 1511) from \emph{Clothing}. Each column is corresponding
		to an aspect attached with an “interpretation” label.}
	\vspace{-0.2cm}
	\begin{tabular}{ccccc} \hline
		\textbf{Value}	&	\textbf{Comfort}	&	\textbf{Accessories}	&	\textbf{Shoes}	&	\textbf{Clothing}	\\ \hline
		price	&	size	&	ring	&	socks	&	shirt	\\
		color	&	fit	&	pretty	&	foot	&	back	\\
		quality	&	wear	&	dress	&	boots	&	bra	\\
		worth	&	comfortable	&	time	&	comfort	&	top	\\
		cute	&	bra	&	beautiful	&	sandals	&	feel	\\
		comfortable	&	small	&	gift	&	walk	&	soft	\\
		fits	&	color	&	earrings	&	toe	&	black	\\
		ring	&	fits	&	compliments	&	pairs	&	jeans	\\
		dress	&	perfect	&	chain	&	hold	&	pants	\\
		shirt	&	material	&	jewelry	&	strap	&	tight	\\
		material	&	long	&	shoes	&	pockets	&	material	\\ \hline
	\end{tabular}
	\label{tab:aspects}
	\vspace{-0.2cm}
\end{table}

\begin{table}[]
	\small
	\centering
	\caption{Interpretation for why the ``user 1511" rated ``item 1" and ``item 2" with 5 and 2, respectively, from \emph{Clothing}.}
	\vspace{-0.2cm}
	\begin{tabular}{c|ccccc} \hline
		Aspects &	Value	&	Comfort	&	Accessories	&	Shoes	&	Clothing	\\ \hline
		Importance (1)	&	0.621	&	0.042	&	0.241	&	0.001	&	0.095	\\ 
		Matching (1)	&	0.982	&	0.596	&	0.660	&	0.759	&	0.638	\\
		Polarity (1)	&	\textbf{\color{red}+}	&	\textbf{\color{blue}-}	&	\textbf{\color{red}+}	&	\textbf{\color{blue}-}	&	\textbf{\color{red}+}	\\ \hline
		Importance (2)	&	0.621	&	0.042	&	0.241	&	0.001	&	0.094	\\
		Matching (2)	&	0.920	&	0.303	&	0.362	&	1.000	&	0.638	\\
		Polarity (2)	&	\textbf{\color{blue}-}	&	\textbf{\color{blue}-}	&	\textbf{\color{blue}-}	&	\textbf{\color{blue}-}		& \textbf{\color{red}+}	\\ \hline
		
	\end{tabular}
	\label{tab:explaination}
	\vspace{-0.2cm}
\end{table}

\subsection{Interpretability (RQ4)} \label{sec:interpret}
In our ALFM model, a user's preference on an item is decomposed into user's preference on different aspects and the importance of those aspects. An aspect is represented as a distribution of latent topics discovered based on reviews. A user's attitude/sentiment on an aspect of the targeted item is decided by the latent factors (learned from ratings) associating with the aspect. Based on the topic distribution of an aspect ($\bm{\theta_{u,a_s}}$) and the word distribution of topics ($\bm{\phi_{w}}$), we can semantically represent an aspect by the top words in this aspect.  Specifically, the probability of a word $w$ in an aspect $a_s$ of a user $u$ can be  computed as $\sum_{k=1}^K\theta_{u,a_s,k}\phi_{k,w}$.  The top 10 aspect words (\#aspect $= 5$) of ``user 1511" from \emph{Clothing} dataset discovered by our model are shown in Table~\ref{tab:aspects}. Notice that in order to obtain a better visualization of each aspect, we removed the “background” words that belong to more than 3 aspects. As shown in Table~\ref{tab:aspects}, the five aspects can be semantically interpreted to ``value"\footnote{``Value" means value for money}, ``comfort", ``accessories", ``shoes", and ``clothing". Next, we illustrate the interpretability of our ALFM model on high or low ratings by examples from the same dataset. Table~\ref{tab:explaination} shows the aspect importance (i.e., $\rho_{u,i,a}$ in Eq.~\ref{eq:rho}) of the ``user 1511" ,  the aspect matching scores (i.e., $s_{u,i,a}$ in Eq.~\ref{eq:jsd}) as well as sentiment polarity (obtained by Eq.~\ref{eq:re2}) on the five aspects with respect to ``item 1" and ``item 2" in \emph{Clothing} dataset. From the results, we can see that ``user 1511" pays more attention to ``Value" and ``Accessories" aspects. On the  ``Value" aspect, both ``item 1" and ``item 2" highly match her preference,  however, she has a positive sentiment on ``item 1" while a negative sentiment on ``item 2".\footnote{As a reminder, the aspect matching is based on the reviews. It is possible that both item 1 and item 2 contains comments on aspect ``value". However,  ``item 1" has a high value while ``item 2" has a low value.} For the  ``Accessories" aspect, ``item 1" has a higher matching score than ``item 2"; and more importantly, the sentiment is positive on ``item 1" while negative on ``item 2". As a result, ``user 1511" rated ``item 1" with 5 while rated ``item 2" with 2. From the examples, we can see that our model could provide explanations for the recommendations in depth with \emph{aspect semantics}, \emph{aspect matching score}, as well as \emph{aspect ratings} (which shows sentiment polarity).

\section{Conclusions} \label{sec:concl}
In this paper, we proposed an aspect-aware latent factor model for rating prediction by effectively combining reviews and ratings. Our model correlates the latent topics learned from review text and the latent factors learned from ratings based on the same set of aspects, which are discovered from textual reviews.  Accordingly, our model does not have the constraint of one-to-one mapping between latent factors and latent topics as previous models (e.g., HFT, RMR, RBLT, etc.), and thus could achieve better user preference and item feature modeling. Besides, our model is able to estimate aspect ratings and assign weights to different aspects. The aspect weight is dependent on each user-item pair, since it is estimated based on user's personal preferences on the corresponding aspect towards an item. Experimental results on 19 real-world datasets show that our model greatly improves the rating prediction accuracy compared to the state-of-the-art methods, especially for users who have few ratings. With the extracted aspects from textual reviews, estimated aspect weights, and aspect ratings, our model could provide interpretation for recommendation results in great detail.

\begin{acks}
This research is supported by the National Research Foundation, Prime Minister's Office, Singapore under its International Research Centre in Singapore Funding Initiative. The authors would like to thank Rose Catherine Kanjirathinkal (from CMU)'s great help on fine-tuning the results of TransNet on all datasets.
\end{acks}

\newpage
\bibliographystyle{ACM-Reference-Format}
\balance
\bibliography{www_long}

\end{document}